\documentclass[aps,prb,twocolumn,superscriptaddress]{revtex4-2} 

\usepackage{graphicx}
\usepackage{amssymb}
\usepackage{amsmath}
\usepackage{amsfonts}
\usepackage{braket,gensymb}
\usepackage{color}
\usepackage[utf8]{inputenc}		
\usepackage[english,russian]{babel}
\usepackage{array}

\begin{document}

\title{ 
On thermal Hall effect in boson systems 
}

\author{Yu.~V. Baramygina}
\affiliation{NRC ``Kurchatov Institute'', Petersburg Nuclear Physics Institute, Gatchina
188300, Russia}
\affiliation{ Ioffe  Institute, Politekhnicheskaya 26,  194021,   St.~Petersburg,   Russia}
\author{D.~N. Aristov}
\affiliation{NRC ``Kurchatov Institute'', Petersburg Nuclear Physics Institute, Gatchina
188300, Russia}
\affiliation{ Ioffe  Institute, Politekhnicheskaya 26,  194021,   St.~Petersburg,   Russia}
\affiliation{St.Petersburg State University, 7/9 Universitetskaya nab., 199034
St.~Petersburg, Russia}

\begin{abstract} 
We theoretically discuss the thermal Hall effect (THE) in boson systems keeping in mind possible applications to magnetically ordered insulators with skyrmion lattice ground state.  
The magnon spectrum of such systems may undergo a topological transition upon the variation of magnetic field, and THE could be a tool to confirm this phenomenon. 
We rederive formulas, suitable for our purposes, and find expressions which have a different temperature dependence of THE 
as compared  to previously reported results. Particularly, in presence of anomalous terms in magnon Hamiltonian the thermal Hall conductivity has a small but finite residual $1/T$ contribution. Such contribution may become sizable at very low temperatures and resembles Meissner effect in superconducting metals. 
\end{abstract}

\maketitle

{\bf Introduction}.
 The topological properties of magnets with non-collinear type of ordering attract a considerable attention over a last decade. \cite{Petrovic2025}  In particular, 
magnetic skyrmions and the ordered arrays of them is  under active discussion.  The ground state of these systems is known as skyrmion crystals (SkX) and possesses non-trivial topological character. The magnon excitations over the ground state of SkX may also be  topologically non-trivial, so that magnon bands are characterized by non zero topological quantities such as Berry curvature and Chern number.
\cite{Nagaosa2013,roldan2016,Weber_2022,Ghader_2024}
 
 It was shown recently that a topological transition in the spectrum of magnons may occur upon the variation of the external magnetic field.
 \cite{Diaz2020,Timofeev_2023c}  
 Such a transition is also sensitive to the exchange anisotropy which may hinder its experimental observation. \cite{Timofeev2026}   
One possibility to check the existence of the topological transition in the spectrum is the measurement of the thermal Hall effect (THE) in these systems \cite{Hoogdalem2013,PhysRevResearch.4.043085}, expecting that the temperature dependence of THE changes at the transition.  
The usual Hall effect in magnetic insulators  is absent  since the low energy magnon excitations are  charge-neutral and THE is determined by the topological characteristic, known as  Berry curvature. 

Looking for appropriate formulas  suitable for modeling THE in boson systems we found an interesting picture, with several papers providing somewhat conflicting results. An earlier paper \cite{Katsura2010} presented a simpler formula for THE conductivity, $\kappa^{xy}$, which however might diverge when $T\to0$. This formula was  disputed in  Refs. \cite{Qin2011,Qin2012,Matsumoto2011,Matsumoto2014}  whose authors pointed out that an additional term, dubbed ``energy magnetization'' or edge current should be added, so that the resulting more complicated form of $\kappa^{xy}$ vanished in the limit $T=0$.  The attempt to reconciliate these formulas in \cite{Mangeolle2024} in a finite width geometry   led to the support of both \cite{Qin2012} and \cite{Matsumoto2014}. The detailed derivation of  $\kappa^{xy}$ was provided in  \cite{Matsumoto2014} and we  took it for our work.

However,  our preliminary study of THE in SkX  in the continuous model  \cite{Timofeev_2023c}  showed that the average Berry curvature, $\Omega^{xy}_{n}$, remained nearly constant upon the increase of  magnon energy, $\varepsilon_n$. This detail together with the particular form of $\kappa^{xy}$ presented in \cite{Matsumoto2014} makes it almost impossible to detect any change of $T$-dependence of THE near the topological transition in the low-energy part of the spectrum. 
Based on these observations, we chose to re-derive the formula for THE, along the lines of Ref. \cite{Matsumoto2014} while paying special attention to  the so-called zero-point fluctuations in our problem.  We did not assume the finite width geometry and edge currents, similarly to what was done in \cite{Matsumoto2014}. Surprisingly, our final expression for $\kappa^{xy}$ differs from one obtained earlier in \cite{Qin2012,Matsumoto2014} and is closer in  its form to one presented in \cite{Katsura2010}. We comment on the significance of our form of $\kappa^{xy}$ for THE in  SkX in the concluding section of our paper.

{\bf Model}.
It was shown earlier \cite{garst2017collective,Timofeev2022}, that the magnon excitations in skyrmion crystals are defined by the energy density 
\begin{equation} h = \frac12 
\begin{pmatrix}
  \bar{\psi},& \psi
\end{pmatrix}
 \hat{\mathcal{H}} 
\begin{pmatrix}
  \psi\\
  \bar{\psi}
\end{pmatrix} \,,
\end{equation} with $\psi$ and its complex conjugate, $\bar{\psi}$, describe the fluctuations of  the local magnetization around its equilibrium value. \cite{Timofeev2022}  The Hamiltonian  of Bogoliubov-de Gennes boson type, $\hat{\mathcal{H}}$, is given by  
\begin{equation}
\hat{\mathcal H} = 
	\begin{pmatrix}
	(-i\nabla + \mathbf A)^2 + U & V \\
	V^* & (i \nabla + \mathbf A)^2 + U
	\end{pmatrix} \,. 
\label{ham}	
\end{equation}
Here $U$, $V$ are periodic potentials, and $\mathbf A$ is vector potential with the zero flux through the unit cell, $\int d{\mathbf r}\,
 \nabla \times \mathbf A =0$. 
The  spectrum, $\varepsilon_{n}$,  can be found from the equation $\hat{\mathcal{H}} \mathrm{\Psi}_n =  \varepsilon_{n} \sigma_3 \mathrm{\Psi}_n $ with $\sigma_3$ Pauli matrix, and the  pseudo-spinor, $\mathrm{\Psi}_n= \begin{pmatrix}  u_n\\  v_n \end{pmatrix}  $. 
In view of the property $\sigma_1\hat{\mathcal{H}}^{\ast}\sigma_1 = \hat{\mathcal{H}} $ the function $\bar{\Psi}_n \equiv \sigma_1 \Psi_n^* = \begin{pmatrix} v_n^*\\  u_n^* \end{pmatrix}  $ also obeys the above spectral equation with $\epsilon_n \to - \epsilon_n$. \cite{schutte2014magnon} It is then convenient to  introduce the  matrix  $\Upsilon_n = \begin{pmatrix}
 u_n&  v_n^*\\
v_n&  u_n^*
\end{pmatrix} $, satisfying  \cite{Timofeev2022}
\begin{equation}
\label{GamUpsilon}
\hat{\mathcal{H}} \Upsilon_n = \varepsilon_n \sigma_3 \Upsilon_n \sigma_3 \,. 
\end{equation}
Notice that the orthogonality relations are 
\begin {equation}
\begin{aligned} 
&
 \int d^2\mathbf{r} \, \Upsilon_n^\dagger 
 \sigma_3 \Upsilon_m = \delta_{nm}  \sigma_3
 \,, \\ 
 &  \sum_n \Upsilon_n (\mathbf{r})
 \sigma_3 \Upsilon_n ^\dagger (\mathbf{r}')  = 
 \delta(\mathbf{r}- \mathbf{r}') \sigma_3 \,.
\end{aligned}
\label{orthog}
\end{equation}  
The second quantization formula reads 
\begin {equation}
\begin{pmatrix}
  \psi \\
  \bar \psi 
\end{pmatrix}  
= \sum_n \Upsilon_n
\begin{pmatrix}
  c_n   \\
   c_n^ \dagger
\end{pmatrix} \,,
\label{psiExpa}
\end{equation} 
with $[c_n,  c_m^ \dagger] = \delta_{nm}$, and as a consequence,  $[ \psi(\mathbf{r}),  \bar \psi (\mathbf{r}')] =  \delta(\mathbf{r}- \mathbf{r}')$, see also \cite{Timofeev_2023a}. Here and below the index, $n$, includes both the wave-vector and the magnon band number.

  {\bf ``Gravitation'' potential}. We assume a small spatial variation of the temperature over the sample, with  
\[
T(\mathbf r) = T\left(1 - \chi(\mathbf r)\right),
\]
and  $\chi(\mathbf{r}) =  \mathbf r \cdot \nabla \chi$. The gradient  $\nabla \chi = - \frac{\nabla T}{T}$ is a small  parameter of our theory.  
This gradient can be represented as the coordinate-dependent ``gravitation'' correction to the Hamiltonian, with the modified statistical  weight factor    
\[
\exp\left(-\tfrac{\hat{\mathcal{H}}}{k_{B} T(\mathbf r)}\right) \simeq \exp\left(-\tfrac{\hat{\mathcal{H}} }{k_{B} T} (1 + \chi (\mathbf r)) \right)\,. 
\]
The total Hamiltonian is then cast in the form 
\[
\hat{\mathcal{H}}_{tot} = \hat{\mathcal{H}} + \hat{F},
\] 
with the symmetrized form of the perturbation    \cite{Luttinger1964}
\begin{equation}
\hat{F} = \tfrac{1}{2} \partial_{\nu}\chi \left(\hat{\mathcal H} r^{\nu} + r^{\nu} \hat{\mathcal H}\right) \,,
\end{equation}
with $\nu=x,y$ and implied summation over Greek indices.  

  {\bf Thermal current}. 
The expression for the heat current, $\mathbf{j}_{E}$, can be found from the continuity equation $\partial_{t} h = -\nabla \mathbf{j}_{E}$, see also \cite{Mangeolle2024}. 
After some calculation presented elsewhere we find 
\begin{equation}
\begin{aligned}
j^{\mu}_{E}  & = \begin{pmatrix}
        \bar \psi, & \psi
        \end{pmatrix}  \hat{ j}^{\mu}_{E} \begin{pmatrix}
        \psi \\
        \bar \psi
        \end{pmatrix} \, , \\ 
 \hat{ j}^{\mu}_{E} &= \tfrac{1}{4} 
 (1+\tfrac{1}{2} \chi  )
 \hat{V}^{\mu}  (1+\tfrac{1}{2} \chi )^2 \sigma_3\hat{\mathcal{H}}  (1+\tfrac{1}{2} \chi ) + H.c.
 \, ,
\end{aligned}    
\label{defCurrent}    
\end{equation}
in accordance with \cite{Matsumoto2014}, 
with the group velocity operator 
\[ \hat{\mathbf{V}} = 
 - i [\mathbf{r}, \mathcal{H}] =
2   \begin{pmatrix} - i\nabla +\mathbf{A} ,& 0   \\ 0, & - i\nabla -\mathbf{A}&   \end{pmatrix} \,. \]  

The first two terms of the Taylor expansion of $ j^{\mu}_{E}$ in  $\partial_{\nu} \chi$ have the form  
\begin{equation}
\begin{aligned}
	\hat{j}_{0}^{\mu} &= \tfrac{1}{4}(\hat{V}^{\mu}\sigma_3\hat{\mathcal{H}}+\hat{\mathcal{H}}\sigma_3\hat{V}^{\mu}) 
	\,, \\
 \hat{j}_{1}^{\mu} &= \tfrac{1}{8}\partial_{\nu}\chi   \left[ 
   4 \hat{\mathcal{H}}\sigma_3 r^{\nu}\hat{V}^{\mu} +  i \hat{V}^{\nu}\sigma_3\hat{V}^{\mu} 
+2i\delta_{\mu\nu} \hat{\mathcal{H}}\sigma_3   + H.c. \right] \,.
\end{aligned}
\label{j1} 
\end{equation}
We note  that the second term of $ \hat{j}_{1}^{\mu}$ when combined with its conjugate gives the curl of the gauge potential, 
$\frac{i}{4}[\hat{V}_{\nu}\sigma_3\hat{V}_{\mu} - \hat{V}_{\mu}\sigma_3 \hat{V}_{\nu}] = \partial_{\nu}A_{\mu} - \partial_{\mu}A_{\nu}$, and  
the third term, $\propto \delta_{\mu\nu}$, does not contribute to thermal Hall current.

{\bf Matrix elements}. 
We define the matrix element of the operator $\hat{O}$ as the 2$\times$2 matrix  of the form 
\begin{equation}
\label{matel}
(\hat{O})_{nm} = \int d\mathbf r\ \Upsilon^{\dagger}_n \hat{O} \Upsilon_m \,, 
\end{equation}
so that $\begin{pmatrix}
        \bar \psi, & \psi
        \end{pmatrix}  \hat{O} \begin{pmatrix}
        \psi \\
        \bar \psi
        \end{pmatrix} =  \sum_{nm}\begin{pmatrix}
  c_n ^ \dagger  , &    c_n
\end{pmatrix} (\hat{O})_{nm} \begin{pmatrix}
  c_m   \\
   c_m^ \dagger
\end{pmatrix}$. 
Using the relation   (\ref{GamUpsilon}), we find the expressions for the matrix elements of the operators  $\hat{j}_{0}^{\mu}$ and $\hat{F}$ ~:  
\begin{equation}
\begin{aligned}
(\hat{j}_{0}^{\mu})_{nm} & = \tfrac{1}{4}\left[ (\hat{V}^{\mu})_{nm}\sigma_3\varepsilon_m + \varepsilon_n\sigma_3(\hat{V}^{\mu})_{nm}\right] \,, \\
(\hat{F})_{nm} &= \tfrac{1}{2}\partial_{\nu}\chi\left[\varepsilon_n \sigma_3 (\sigma_3 r^{\nu})_{nm} + (\sigma_3r^{\nu})_{nm} \varepsilon_m \sigma_3\right] \,.
\end{aligned}
\label{j0matel}
\end{equation}
Here and below we repeatedly use the rule $(\hat{A} \hat{B} )_{nm}  = \sum_{k} (\hat{A}   )_{nk}\sigma_{3}  ( \sigma_{3} \hat{B} )_{km}
= \sum_{k} (\hat{A}  \sigma_{3} )_{nk}\sigma_{3}  (  \hat{B} )_{km}$ following from Eq.\ \eqref{orthog}. 
We now want to express 
$(\hat{F})_{nm}$ entirely in terms of the velocity operator,  $(\hat{V}^{\mu})_{nm}$. We have 
\[
\label{matelV}
\begin{aligned}
(\hat{V}^{\nu})_{nm} & =  i  (\hat{\mathcal{H}} r^{\nu} -  r^{\nu}\hat{\mathcal{H}} )_{nm} \\
& = i\left(\varepsilon_n \sigma_3 (\sigma_3 r^{\nu})_{nm} - (\sigma_3 r^{\nu})_{nm} \varepsilon_m \sigma_3\right).
\end{aligned}
\]
Introducing  $\eta = (\hat{V}^{\nu})_{nm}$ and $\xi = (\sigma_3 r^{\nu})_{nm}$ we observe that 
\begin{equation*}
    \begin{pmatrix}
	\eta \\
	\sigma_3 \eta \sigma_3
    \end{pmatrix} = i
    \begin{pmatrix}
	\varepsilon_n & -\varepsilon_m \\
	-\varepsilon_m & \varepsilon_n
    \end{pmatrix}
    \begin{pmatrix}
	\sigma_3 \xi \\
	\xi \sigma_3
    \end{pmatrix} \, .
\end{equation*} 
Resolving this equation with respect to $\xi$, we find 
\begin{equation}
(\sigma_3 r^{\nu})_{nm} = 
 \frac{i}{\varepsilon_m^2 - \varepsilon_n^2}
 \left(
 \varepsilon_n \sigma_3 (\hat{V}^{\nu})_{nm} + \varepsilon_m (\hat{V}^{\nu})_{nm} \sigma_3 \right) \,, 
 \label{sigmarmunu}
\end{equation} 
and the desired expression is 
\begin{equation}
\begin{aligned}
& (\hat{F})_{nm} = - \frac{i}{2}\frac{\partial_{\nu}\chi}{\varepsilon_n^2 - \varepsilon_m^2} \\
& \times\left[(\varepsilon_n^2 + \varepsilon_m^2) (\hat{V}^{\nu})_{nm} + 2\varepsilon_n\varepsilon_m \sigma_3 (\hat{V}^{\nu})_{nm} \sigma_3\right] \,, 
\end{aligned}
\label{Fmatel}
\end{equation}
we note that $F_{nm}^{11} = F_{mn}^{22}$, 
$F_{nm}^{12} = F_{mn}^{12}$, $F_{nm}^{21} = F_{mn}^{21}$.

Let us discuss the matrix elements of the velocity operator,  $(\hat{V}^{\nu})_{nm}$. It is easy to show that there are only two independent complex numbers there, and one can represent $(\hat{V}^{\nu})_{nm}$ in the form 
\begin{equation}
(\hat{V}^{\mu})_{nm} =
    \begin{pmatrix} 
    V^{\mu}_{a,nm} & -(V^{\mu}_{b,nm})^* \\
    V^{\mu}_{b,nm} & -(V^{\mu}_{a,nm})^*
    \end{pmatrix} \,,
\label{Vmatel}
\end{equation}
where $V^{\mu}_{a,nm} = \int d\mathbf{r}\, \mathrm{\Psi}_n^{\dagger}\hat{V}^{\mu}\mathrm{\Psi}_m$, $V^{\mu}_{b,nm} = \int d\mathbf{r}\, \bar{\mathrm{\Psi}}_n^{\dagger}\hat{V}^{\mu}\mathrm{\Psi}_m$, with the above $ {\Psi}_n$, $\bar{ {\Psi}}_n$. 

It is worth also to discuss here the matrix elements $( r^{\nu})_{mn}$,    appearing in our analysis below.   Repeating the  above steps, which led us from $( \sigma_{3} r^{\nu})_{mn}$ to $(\hat{V}^{\nu})_{mn}$, 
we similarly derive 
 \begin{equation} 
\begin{aligned}
( r^{\nu})_{mn} & =   \frac{i}{\varepsilon_{n}^{2}-\varepsilon_{m}^{2}}    
\left[  \varepsilon_{m} \sigma_3 {  (\widetilde{V}^{\nu})_{mn} }
+ \varepsilon_{n}  { (\widetilde{V}^{\nu})_{mn} }\sigma_3 \right] 
 \,,
\end{aligned}
\label{rmunu}
\end{equation}  
{  where $\widetilde{V}^{\nu} = - i [\sigma_3 r^{\nu}, \mathcal{H}]$ and} the matrix
 \begin{equation} 
\begin{aligned} 
(\widetilde{V}^{\nu} )_{nm}   & = 
  \begin{pmatrix} \widetilde{V}^{\nu}_{a,nm}  , & -(\widetilde{V}^{\nu}_{b,nm})^{\ast}  \\  - \widetilde{V}^{\nu}_{b,nm} , & (\widetilde{V}^{\nu}_{a,nm})^{\ast}   \end{pmatrix}
\,.
\end{aligned}
\label{sigmaVmunu}
\end{equation} 
One can check that the  2$\times$2 matrix  $( \hat{ V} )_{mn} $ is the Hermitian conjugate to $( \hat{ V} )_{nm} $
and  $(\widetilde{V}^{\nu})_{mn}  = [(\widetilde{V}^{\nu})_{nm}]^{\dagger}$, as well. It means 
that      $ {V}^{\nu}_{a,mn} =  ({V}^{\nu}_{a,nm})^{\ast}$, $ {V}^{\nu}_{b,mn} =  -{V}^{\nu}_{b,nm}$ and 
   $ \widetilde{V}^{\nu}_{a,mn} =  (\widetilde{V}^{\nu}_{a,nm})^{\ast}$, $ \widetilde{V}^{\nu}_{b,mn} =  \widetilde{V}^{\nu}_{b,nm}$. 

In what follows, we use also bilinear in $( \hat{ V} )_{mn} $ combinations. We define here 
\begin{equation}
\begin{aligned}
X^{\mu\nu}_{c,nm} & = i (V_{c,nm}^{\mu}(V_{c,nm}^{\nu})^* - (V_{c,nm}^{\mu})^* V_{c,nm}^{\nu} )\,,
\\ 
\widetilde{X}^{\mu\nu}_{c,nm} &=i( V_{c,nm}^{\mu}(\widetilde{ V}_{c,nm}^{\nu})^* 
- (V_{c,nm}^{\mu})^* \widetilde{ V}_{c,nm}^{\nu})\,,
\end{aligned}
\label{X}
\end{equation} 
with $c = a,b $. 
We observe that $X_{a,nm}^{\mu\nu} = - X_{a,mn}^{\mu\nu}$,  $X_{b,nm}^{\mu\nu} = X_{b,mn}^{\mu\nu}$ and 
$\widetilde{X}_{a,nm}^{\mu\nu} = - \widetilde{X}_{a,mn}^{\mu\nu}$,  $\widetilde{X}_{b,nm}^{\mu\nu} = -\widetilde{X}_{b,mn}^{\mu\nu}$.

 {\bf  Linear response theory}. We  assume that the perturbation $\hat{F}$ is adiabatically switched on in time, $\hat{F}e^{\delta t}$, with $\delta \to +0$. The standard formula for the linear-in-$\chi(\mathbf{r})$ response  gives 
\begin{equation}
\label{linresponse}
\langle \hat{j}^{\mu} \rangle = \langle \hat{j}^{\mu} \rangle_{0} - i\int_{-\infty}^{0} dt\, \langle [\hat{j}^{\mu}(0), \hat{F}(t)]\rangle _{0}
\, e^{\delta t}\,,
\end{equation}
where in the interaction picture $\hat{A}(t) = e^{i\hat{\mathcal{H}} t}\hat{A}e^{-i\hat{\mathcal{H}} t} $ and the averaging with the unperturbed Hamiltonian reads 
$\langle \hat{A}\rangle_{0}= \mbox{Tr}(e^{-\hat{\mathcal{H}} /k T }\hat{A} ) /\mbox{Tr}(e^{-\hat{\mathcal{H}} /k T } )$.
The expression $ \hat{j}_0^{\mu} $ is independent of $\chi(\mathbf{r})$ and vanishes upon averaging, $\langle\hat{j}_0^{\mu}\rangle_0 = 0$. Following conventional notation, we write 
$\langle\hat{j}_1^{\mu}\rangle_{0} \equiv -M^{\mu\nu} \partial_{\nu}\chi$, and $ - i\int_{-\infty}^{0} dt\, \langle [\hat{j}_0^{\mu}(0), \hat{F}(t)]\rangle_{0} \equiv -S^{\mu\nu}\partial_{\nu}\chi$ so that the linear-in-$\partial_{\nu}\chi$ contribution takes the form
\[
\langle \hat{j}^{\mu} \rangle = - (S^{\mu\nu}+ M^{\mu\nu})\partial_{\nu}\chi,
\]
and the heat conductivity tensor is  
\begin{equation}
\kappa^{\mu\nu} = \tfrac{1}{T} (S^{\mu\nu}+ M^{\mu\nu})\,.
\label{defkappa}
\end{equation}
We note that the terms $S^{\mu\nu}$  and $M^{\mu\nu}$ are similar to paramagnetic and diamagnetic terms, respectively, in the theory of electrical conductivity of metals. 

First we discuss the coefficient $S^{\mu\nu}$. Using properties of the matrix elements of $\hat{F}$,  $\hat{j}_0^{\mu}$, we find  \begin{equation*}
\begin{aligned}
S^{\mu\nu} = &-\frac{1}{2}\sum_{n,m}\bigg[ \frac{2N(\varepsilon_n)}{\varepsilon_n - \varepsilon_m} 
\left( (j^{\mu}_{0})_{nm}^{11}(F)_{mn}^{11} + (j^{\mu}_{0})_{nm}^{22}(F)_{mn}^{22}\right)\\
&-\frac{2N(\varepsilon_n)+1}{\varepsilon_n + \varepsilon_m}
\left((j^{\mu}_{0})_{nm}^{12}(F)_{mn}^{21} +(j^{\mu}_{0})_{nm}^{21}(F)_{mn}^{12}\right)\bigg]\,.
\end{aligned}
\end{equation*}
Making use of Eqs.\ \eqref{j0matel}, \eqref{Fmatel} and \eqref{Vmatel}, we come to a compact expression 
\begin{equation}
\begin{aligned}
S^{\mu\nu} = -\tfrac{1}{8}\sum_{n,m} \Bigg[ &2N(\varepsilon_n)\frac{(\varepsilon_n+\varepsilon_m)^2}{(\varepsilon_n - \varepsilon_m)^2}X^{\mu\nu}_{a,nm} \\
& + (2N(\varepsilon_n)+1)\frac{(\varepsilon_n - \varepsilon_m)^2}{(\varepsilon_n + \varepsilon_m)^2}X^{\mu\nu}_{b,nm}\Bigg] \,,
\end{aligned}
\label{Smunu}
\end{equation} 
with $X^{\mu\nu}_{c,nm}$ defined in Eq. \eqref{X}. 
Deriving the above expression for $S^{\mu\nu} $, we focussed on the thermal Hall current, implying off-diagonal tensor structure, $\mu\neq \nu$. 
It allowed us to ignore the intra-band transitions, $\varepsilon_n = \varepsilon_m$,  in \eqref{Smunu};  otherwise we would need to resolve the appearing singularities by considering finite-lifetime effects. \cite{Katsura2010}

It is convenient to represent the coefficient $M_{\mu\nu}$ by the sum 
\[
M^{\mu\nu} = M^{\mu\nu}_1 + M^{\mu\nu}_2 \,, 
\]
with $M^{\mu\nu}_1$,  $M^{\mu\nu}_2$ stemming from two first terms in $\hat{j}_{1}^{\mu}$, Eq.\ \eqref{j1}, plus their Hermitian conjugates, respectively. 
\begin{equation}
\begin{aligned}
 M^{\mu\nu}_1 &= -\tfrac{1}{2}\langle\hat{\mathcal{H}}\sigma_3 r^{\nu}\hat{V}^{\mu} + \hat{V}^{\mu}r^{\nu}\sigma_3\hat{\mathcal{H}}\rangle_0 \\
& = -\tfrac{1}{2}\sum_{n,m}\varepsilon_n\left(2N(\varepsilon_n)+1\right)\left[\frac{X_{a,nm}^{\mu\nu}}{\varepsilon_n - \varepsilon_m} + \frac{X_{b,nm}^{\mu\nu}}{\varepsilon_n+\varepsilon_m}\right] \, , \\ 
 M^{\mu\nu}_2 &= \tfrac{i}{8}\langle \hat{V}^{\mu}\sigma_3 \hat{V}^{\nu} - \hat{V}^{\nu}\sigma_3\hat{V}^{\mu}\rangle_{0} \\
& = \tfrac{1}{8}\sum_{n,m}\left(2N(\varepsilon_n)+1\right)\left[X_{a,nm}^{\mu\nu} + X_{b,nm}^{\mu\nu}\right] \, .
\end{aligned}
\label{coefM}
\end{equation}
Combining the contributions \eqref{Smunu}, \eqref{coefM} we arrive to the expression 
\begin{equation}
\kappa^{\mu\nu} = -\frac{1}{T} \sum_n \varepsilon_n^2 \left(N(\varepsilon_n)+\frac{1}{2}\right) \Omega_n^{\mu\nu}\,,
\label{kappa1}
\end{equation}
where the Berry curvature $\Omega_n$ is defined by 
\begin{equation}
\Omega_n^{\mu\nu} =  \sum_m \left[\frac{X_{a,nm}^{\mu\nu}}{(\varepsilon_n - \varepsilon_m)^{2}} + \frac{X_{b,nm}^{\mu\nu}}{(\varepsilon_n + \varepsilon_m)^2}\right] \,.
\label{Berry1}
\end{equation}
Since the Bose distribution function $N(\varepsilon_n)$ in the first term in \eqref{kappa1}  essentially restricts  the summation over $n$ by  $\varepsilon_n < T$, the corresponding contribution to $\kappa^{\mu\nu}$ is bounded from above. 
On the other hand, $T$-independent second term  in the sum \eqref{kappa1} leads at $T\to 0$  to the divergent contribution $\kappa^{\mu\nu} \sim 1/T$ which is a loose analog  of the Meissner effect in superconductivity.

Below we show  that the divergent at $T\to 0$ contribution in \eqref{kappa1} is overestimated and the correct procedure reduces its value, although the divergence of  $\kappa^{\mu\nu} $  \emph{ does not completely disappear} from the theory.

%
%

{\bf  Zero-point fluctuations}.
Our key observation is the existence of so-called zero-point fluctuations in the current  $\hat{j}_{1}^{\mu}$, which should be analyzed together with  Eq. \eqref{j1}. 
{  In order to see it, let us set $V\equiv 0$ in Eq.\ \eqref{ham}, it follows then that  $v_{n}= 0$ in  $\Upsilon_n$ and  $X_{b,nm}^{\mu\nu}   =0$.  In view of antisymmetry of $X_{a,nm}^{\mu\nu}$  in indices $n$, $m$ we have  $S^{\mu\nu}= M^{\mu\nu}_2 = 0$ while $M^{\mu\nu}_1 \neq  0$ in the limit $T=0$, thus leading to divergence in \eqref{kappa1}.  Clearly in the absence of anomalous terms  $V $ in Eq.\ \eqref{ham} the expression for current $\hat j_0^\mu$ should have the usual form  $\varepsilon_n (\partial \varepsilon_n /\partial q)\,   c_n^\dagger c_n$ rather than the form $\tfrac 12 \varepsilon_n (\partial \varepsilon_n /\partial q)\,  ( c_n^\dagger c_n +   c_n  c_n^\dagger) $ stemming from Eq. \eqref{j1}; the same is expected for $\hat j_1^\mu$.  As a result the average   current  $\hat j_1^\mu$ should contain in this case the expression $\langle  c_n^\dagger c_n \rangle = N(\varepsilon_n)$  instead of $\tfrac 12 \langle  c_n^\dagger c_n +   c_n  c_n^\dagger \rangle = N(\varepsilon_n)+1/2$ in Eq. \eqref{coefM}. In case of  $V\neq  0$ the situation is more complicated and discussed below. 

In general, } let us consider a normal-ordered quadratic-in-$\psi$ operator, $\hat A$, corresponding to observable quantity, 
$\hat A =  \bar \psi A^{11}\psi + \tfrac12  \bar \psi A^{12}\bar \psi + \tfrac12   \psi A^{21}\psi$.  
A self-conjugate character of such observable operator dictates that $A^{11} = (A^{11})^{\dagger} $, $A^{12} = (A^{21})^{\ast} $. For our convenience, we pass from this normal-ordered form  to the expression  
  \begin{equation}
  \label{normalA}
\begin{aligned}
\hat A & = \tfrac12  \begin{pmatrix} \bar \psi, &   \psi \end{pmatrix} 
  \begin{pmatrix} A^{11} , & (A^{21})^{\ast} \\   A^{21} , & (A^{11})^{\ast}    \end{pmatrix}
   \begin{pmatrix} \psi \\ \bar \psi \end{pmatrix} 
    \\ &   + \tfrac12
   \begin{pmatrix} \bar \psi, &   \psi \end{pmatrix} 
  \begin{pmatrix} A^{11} , & 0 \\   0 , & -(A^{11})^{\ast} 
  \end{pmatrix} 
 \begin{pmatrix} \psi \\ \bar \psi \end{pmatrix} 
\,.
\end{aligned}
\end{equation}
Here the first term is used for further matrix operations, and   
the second term is proportional to the commutator, $ [ \psi, \bar \psi] $, 
which is $c$-number and may usually be ignored. 

{  The} latter term becomes important, e.g., in the discussion of  difference between the N\'eel and quantum N\'eel states of antiferromagnet.  
In this case the observable $\hat A$ is the Hamiltonian, with $A^{11}= J(0)$ and $A^{12} = J(\mathbf{q})$, here $J(\mathbf{q})$ is the Fourier transform of antiferromagnetic exchange.   After diagonalization this Hamiltonian  acquires the form 
$\tfrac12 \sum_{q}\varepsilon(\mathbf{q})(c^{\dagger}_{\mathbf{q}} c _{\mathbf{q}}+c_{\mathbf{q}} c^{\dagger} _{\mathbf{q}} )$,  with $\varepsilon(\mathbf{q}) = \sqrt{J(0)^{2} - J(\mathbf{q})^{2} }$. 
Reducing it again to normal-ordered form,  $H = \sum_{q}\varepsilon(\mathbf{q}) c^{\dagger}_{\mathbf{q}} c _{\mathbf{q}}$, we obtain another $c$-number. The difference of these two  $c$-numbers, $\Delta E = \tfrac12 \sum_{q} (\varepsilon(\mathbf{q}) - J(0))$, gives the energy gain of quantum N\'eel state over the naive N\'eel ordered state. 
  
{  The same reasoning may be applied}
to the discussion of the current,  $\langle \hat{j}_{1}^{\mu}\rangle$, which in addition to already calculated contributions \eqref{coefM} should contain the part omitted in \eqref{j1} but contained in the true initial form, normally ordered in $\bar \psi, \psi$.  Notice that 
in the semiclassical approach \cite{Timofeev_2023a} the quantities $\bar \psi, \psi$ become operators only after procedure of second quantization, and the question of normal ordering may not seem evident. 
 At the same time,  
 if one  
 uses Maleyev-Dyson representation of spin operators 
 aligned with the local magnetization, \cite{schutte2014magnon,roldan2016,Diaz2020} then the normal ordering of observable expressions becomes rather obvious.

{\bf  Correction term to current $\hat{j}_{1}^{\mu}$}.
Inspecting Eq.\ \eqref{normalA} we note that the missing zero-point contribution, i.e.\  the second term in \eqref{normalA}, can be restored  as  
\[ \hat{R}^{\mu} = \tfrac12 (\sigma_{3}\hat{j}_{1}^{\mu} + \hat{j}_{1}^{\mu}\sigma_{3}) \,. \]  
After some algebra, we can represent it in the form  
\[
\begin{aligned}
\hat{R}^{\mu} & = 
 \tfrac 14   \left(
  \hat{V}^{\mu}  r^{\nu}  \hat{\mathcal H}  + 
\hat{V}^{\mu} \sigma_3 \hat{\mathcal H} r^{\nu} \sigma_3  
  +\tfrac i2 \hat{V}^{\mu} \hat{V}^{\nu}
  \right) \partial_{\nu}\chi + H.c.
\,, 
\end{aligned}
\]
and the average reads 
\begin{equation} 
\begin{aligned}
\langle \hat{R}^{\mu} \rangle & =  
  \sum_{n} \left((N(\varepsilon_{n})+\tfrac12) \mbox{Tr} (\hat{R}^{\mu})_{nn}   - 
\tfrac12 \mbox{Tr}  (\sigma_3 (\hat{R}^{\mu})_{nn} ) \right)
\,. 
\end{aligned}
\label{averageR}
\end{equation}

The diagonal matrix element is given by 
 \begin{equation} 
\begin{aligned}
(\hat{R}^{\mu})_{nn}& =   \tfrac 14  \sum_{m}  
(\hat{V}^{\mu})_{nm} \Big( 
\sigma_3  ( r^{\nu})_{mn} \sigma_3   \varepsilon_{n} 
 \\ & + \varepsilon_{m} ( r^{\nu})_{mn}  +\tfrac i2  \sigma_3 (\sigma_3\hat{V}^{\nu})_{mn}
 \Big) \partial_{\nu}\chi 
   + H.c. 
\,. 
\end{aligned}
\end{equation} 
 Using the properties of $( r^{\nu})_{mn}$, $(\widetilde{V}^{\nu})_{mn}$, $(\sigma_3\hat{V}^{\nu})_{mn}$, Eqs.\ \eqref{rmunu}, \eqref{sigmaVmunu},
we can verify that $\mbox{Tr} (\hat{R}^{\mu})_{nn} =0$ in Eq.\ \eqref{averageR}, 
which shows that $\langle \hat{R}^{\mu} \rangle $ is independent of $T$, as expected.  
The second term, $ - 
\tfrac12
  \sum_{n}      \mbox{Tr}  (\sigma_3 (\hat{R}^{\mu})_{nn} )$, is non-zero and we obtain  
 \begin{equation*} 
\begin{aligned}
\langle \hat{R}^{\mu} \rangle 
& =   - \tfrac14
  \sum_{nm}      \left( 
 \frac{\varepsilon_{n} +\varepsilon_{m} }{\varepsilon_{n} -\varepsilon_{m} }
 \widetilde{X}^{\mu\nu}_{a,nm}  +   
\frac{\varepsilon_{n} -\varepsilon_{m} }{\varepsilon_{n} +\varepsilon_{m} } 
 \widetilde{X}^{\mu\nu}_{b,nm}
   \right) 
   \partial_{\nu}\chi 
 \, .  
\end{aligned}
\end{equation*}  
Using the properties of  $\widetilde{X}^{\mu\nu}_{\xi,nm}$,   Eq. \eqref{X}, we can represent the zero-point contribution as 
 \begin{equation} 
\begin{aligned}
\langle \hat{R}^{\mu} \rangle  
& = 
 -\tfrac 12 \sum_{n} 
 \varepsilon_{n}^{2} \,  \widetilde{\Omega}_n^{\mu\nu} \partial_{\nu}\chi 
 \,,   
\end{aligned}
\label{averageR1}
\end{equation} 
with the modified Berry curvature 
\begin{equation}
 \widetilde{\Omega}_n^{\mu\nu}   = 
\sum_m \left[\frac{\widetilde{X}_{a,nm}^{\mu\nu}}{(\varepsilon_n - \varepsilon_m)^{2}} + \frac{\widetilde{X}_{b,nm}^{\mu\nu}}{(\varepsilon_n + \varepsilon_m)^2}\right] \,. 
\end{equation}

The total expression for the thermal Hall  conductivity acquires the form 
\begin{equation}
\kappa^{\mu\nu} = -\frac{1}{T} \sum_n \varepsilon_n^2 \left[N(\varepsilon_n)\,\Omega_n^{\mu\nu} +
\tfrac{1}{2} \left( \Omega_n^{\mu\nu} - \widetilde{\Omega}_n^{\mu\nu}  \right) \right] \,.
\label{kappa2}
\end{equation}
One can verify that the tensor $\kappa^{\mu\nu}$ is antisymmetric in indices $\mu$, $\nu$. 
Eq. \eqref{kappa2} is the central result of this paper and we discuss it below.

{\bf Discussion and conclusions}. 
Firstly, we note that  in the absence of anomalous terms of the Hamiltonian, $V\equiv0$ in Eq.\ \eqref{ham}, we have $v_{n}\equiv 0$ in  $\Upsilon_n$ and therefore $ V^{\nu}_{a,nm} = \widetilde{V}^{\nu}_{a,nm}$, $ V^{\nu}_{b,nm} = \widetilde{V}^{\nu}_{b,nm}=0 $ in \eqref{Vmatel}, \eqref{sigmaVmunu}. 
It follows   that $X_{a,nm}^{\mu\nu}  = \widetilde{X}_{a,nm}^{\mu\nu}$, $X_{b,nm}^{\mu\nu}  = \widetilde{X}_{b,nm}^{\mu\nu}=0$ and 
$\Omega_n^{\mu\nu} = \widetilde{\Omega}_n^{\mu\nu} $.
In this case the second term in parenthesis in\eqref{kappa2} vanishes and our result simplifies to 
\begin{equation} \kappa^{\mu\nu} = -\frac{1}{T} \sum_n \varepsilon_n^2 \ N(\varepsilon_n)\,\Omega_n^{\mu\nu} \,.
\label{kappa0}
\end{equation}
This expression vanishes in the limit $T\to0$, which makes it different from 
 Ref.\ \cite{Katsura2010} where the terms $M^{\mu\nu}$ were omitted and whose Eq.(3)  corresponds to setting  $\kappa^{\mu\nu}= S^{\mu\nu}/T$ with our Eq.\ \eqref{Smunu}.
 
The terms $M^{\mu\nu}$ were taken into account in Refs.\ \cite{Qin2012,Matsumoto2014} which led to vanishing value of $\kappa^{\mu\nu}$ at $T\to0$.  
However, comparing Eq. (20) in Ref.\ \cite{Qin2012} even with our simpler Eq. \eqref{kappa0} shows a big difference in the  functional form of these  formulas.  
The recent paper \cite{Mangeolle2024} reported an agreement with both \cite{Qin2012} and \cite{Matsumoto2014},  so we focus on the  latter work.  

Having  the same expression for the current \eqref{defCurrent}, 
we observe that some intermediate expressions obtained  in \cite{Matsumoto2014} and in our work coincide. Namely, we agree in formulas for $S^{\mu\nu}$, $ M^{\mu\nu}_2$ and disagree in $ M^{\mu\nu}_1$, our Eqs. \eqref{Smunu}, \eqref{coefM}. 
The contribution of the zero-point fluctuations was not discussed  in \cite{Matsumoto2014}. 
Accordingly,  the explicit analytical forms of our result, 
Eq.\ \eqref{kappa2},  and that of Eq. (32) in \cite{Matsumoto2014} are different.
{ 
We provide the details of our calculation of Eq.\  \eqref{coefM} in the Supplementary material to this paper. The possible source of difference between our results may be suggested as follows. The setup of the problem was explicitly outlined as the  four- terminal geometry in \cite{Matsumoto2011,Mangeolle2024}, with the topological material being placed between the terminals maintained at given temperatures.  The 
presence of confining potential for magnons at the edge of the material and the corresponding edge current were essential for the derivation there.
In our approach of linear response theory we do not assume confining potential for magnons and the edge current does not appear. It would be interesting to resolve this issue rigorously from the Keldysh formalism. \cite{Kamenev2011,Aristov2014}
}

Now let us discuss the significance of two terms of the sum \eqref{kappa2} for the particular case of  thermal Hall effect in 2D skyrmion crystals.  
Preliminary estimates show that $\Omega_n^{\mu\nu}$ roughly remains constant with the increase of  $\varepsilon_n$. 
The difference $(\Omega_n^{\mu\nu}- \widetilde{\Omega}_n^{\mu\nu})$, on the other hand, scales as  $\varepsilon_0^{2}/\varepsilon_n^{2}$ in the wide range $\varepsilon_0\alt \varepsilon_n \alt J$, with $\varepsilon_0 = D^{2}/J \ll J $, here $J$ and $D$  are exchange constant and Dzyaloshinskii-Moriya interaction, respectively.  It follows that the first contribution   to  $\kappa^{\mu\nu}$ is restricted by the energies $\sim T$ and behaves as $ \propto T^{2}/J$ and the second is mildly divergent at high energies and behaves as $\propto \varepsilon_0^{2} /T$. The zero-point fluctuation contribution becomes dominant only at low temperatures,  $T \alt J \cdot(\varepsilon_0/J)^{2/3}$, which may not exceed 1K in realistic compounds.  We plan to further analyze it  elsewhere.

In conclusion, we revisit the formula for the thermal Hall effect in boson systems,  with possible application to non-collinear magnets with topological character of magnon dispersion. We show the existence of residual contribution of zero-point fluctuations, which leads to divergent thermal Hall conductivity in the limit $T\to 0$. 
 
{\bf Acknowledgments}. We thank V.E. Timofeev for numerous helpful discussions. 
This work was supported by the Russian Science Foundation (project no. 25-12-00212), https://rscf.ru/en/project/25-12-00212/. 
The work was partially supported by the Foundation for the Advancement of Theoretical Physics BASIS.


%

\vfill

\newpage

\begin{widetext}
{ 
\begin{center} {Supplementary material} \end{center}
 
In this part we provide a detailed derivation of our result for the quantity $M^{\mu\nu}_1$ in \eqref{coefM}. 
We note that our index $n$ in Eq.\  \eqref{GamUpsilon} corresponds to both  the wave-vector in the Brillouin zone and the spin wave band number $n$ in notation of \cite{Matsumoto2014}.  The identification of our $\Upsilon_n$ with the matrix $T_{\mathbf{k}}$ shows the equivalence of the first line in our \eqref{coefM} with Eq. (D3) in \cite{Matsumoto2014}.  

We consider the quantity 
\begin{equation}
M^{\mu\nu}_1 = -\tfrac{1}{2}\langle\hat{\mathcal H}\sigma_3 r^{\nu}\hat{V}^{\mu}+\hat{V}^{\mu}r^{\nu}\sigma_3\hat{\mathcal H}\rangle_{0} \,, 
\end{equation}
whose average may be written as 
\begin{equation}
\begin{aligned}
M^{\mu\nu}_1 & = -\frac{1}{2} \sum_n \mathrm{Tr}\left[   
\begin{pmatrix} 
N(\varepsilon_n) , & 0 \\ 0 , & N(\varepsilon_n)+1
\end{pmatrix}\cdot
(\hat{\mathcal H}\sigma_3 r^{\nu}\hat{V}^{\mu}+\hat{V}^{\mu}r^{\nu}\sigma_3\hat{\mathcal H} ) _{nn} 
 \right] 
 \\ 
 & = -\frac{1}{2} \sum_n \mathrm{Tr}\left[   \varepsilon_n
\begin{pmatrix} 
N(\varepsilon_n) , & 0 \\ 0 , & N(-\varepsilon_n) 
\end{pmatrix}\cdot
(  r^{\nu}\hat{V}^{\mu}+\hat{V}^{\mu}r^{\nu}  ) _{nn} 
 \right]
  \,.
 \end{aligned}
\end{equation}
The 2$\times$2 matrix $(\ldots)_{nn}$ here may be represented as follows
\begin{equation}
\begin{aligned}
& {\cal C}_{nn} \equiv \left(\hat{\mathcal H}\sigma_3 r^{\nu}\hat{V}^{\mu}+\hat{V}^{\mu}r^{\nu}\sigma_3\hat{\mathcal H}\right)_{nn} = \int d\mathbf r\  \Upsilon^{\dagger}_n \left(\hat{\mathcal H}\sigma_3 r^{\nu}\hat{V}^{\mu}+\hat{V}^{\mu}r^{\nu}\sigma_3\hat{\mathcal H}\right) \Upsilon_n \\
& = \varepsilon_n\sum_m \left[\sigma_3\int d\mathbf r \left(\Upsilon_n^{\dagger}r^{\nu}\sigma_3\Upsilon_m\right)\sigma_3\int d\mathbf r'\left(\Upsilon_m^{\dagger}\hat{V}^{\mu}\Upsilon_n\right) + \int d\mathbf r' \left(\Upsilon_n^{\dagger} \hat{V}^{\mu}\Upsilon_m\right)\sigma_3\int d\mathbf r \left(\Upsilon_m^{\dagger}\sigma_3 r^{\nu}\Upsilon_n\right) \sigma_3\right] \\
& = \varepsilon_n \sum_m \left[\sigma_3 \left(\sigma_3r^{\nu}\right)_{nm}\sigma_3(\hat{V}^{\mu})_{mn} + (\hat{V}^{\mu})_{nm}\sigma_3\left(\sigma_3 r^{\nu}\right)_{mn}\sigma_3\right]
\end{aligned}
\label{M1nn}
\end{equation}
Now using Eq.\  \eqref{sigmarmunu} we express
 $(\sigma_3 r^{\nu})_{nm}$ in terms of the matrix elements of $\hat{V}^{\nu}$
\begin{equation}
\begin{aligned}
\sigma_3(\sigma_3 r^{\nu})_{nm}\sigma_3 &= \frac {-i}{\varepsilon_n^2 - \varepsilon_m^2}\left[\varepsilon_n (\hat{V}^{\nu})_{nm}\sigma_3 + \sigma_3 (\hat{V}^{\nu})_{nm}\varepsilon_m\right] \,, \\ 
\sigma_3 (\sigma_3r^{\nu})_{mn}\sigma_3 & = \frac{i}{\varepsilon_n^2 - \varepsilon_m^2}\left[\varepsilon_n\, \sigma_3 (\hat{V}^{\nu})_{mn} + \varepsilon_m\, (\hat{V}^{\nu})_{mn}\sigma_3\right] \,,
\end{aligned}
\end{equation}
so that   Eq. (\ref{M1nn})  becomes 
\begin{equation}
\begin{aligned}
{\cal C}_{nn}  = 
&  \sum_m \frac{ i \,\varepsilon_n}{\varepsilon_n^2 - \varepsilon_m^2} \left[
-\big(\varepsilon_n (\hat{V}^{\nu})_{nm}\sigma_3 + \sigma_3 (\hat{V}^{\nu})_{nm}\varepsilon_m\big)
(\hat{V}^{\mu})_{mn} + 
(\hat{V}^{\mu})_{nm} 
\big(\varepsilon_n\, \sigma_3 (\hat{V}^{\nu})_{mn} + \varepsilon_m\, (\hat{V}^{\nu})_{mn}\sigma_3\big)
\right] \\
= & \sum_m \frac{i\, \varepsilon_n}{\varepsilon_n^2 - \varepsilon_m^2} \left[\varepsilon_n
\big((\hat{V}^{\mu})_{nm}\sigma_3(\hat{V}^{\nu})_{mn} - (\hat{V}^{\nu})_{nm}\sigma_3(\hat{V}^{\mu})_{mn}\big) 
+ \varepsilon_m
\big(\sigma_3(\hat{V}^{\mu})_{nm}(\hat{V}^{\nu})_{mn} - (\hat{V}^{\nu})_{nm}(\hat{V}^{\mu})_{mn}\sigma_3\big)
\right].
\end{aligned}
\end{equation}
Using a formula similar to (\ref{averageR}),  we represent   $\begin{pmatrix} 
N(\varepsilon_n) , & 0 \\ 0 , & N(\varepsilon_n)+1
\end{pmatrix}  = (N(\varepsilon_n)+\tfrac{1}{2}) \sigma_{0} - \tfrac{1}{2} \sigma_{3}$. The structure of matrices $(\hat{V}^{\mu})_{nm}$, Eq.  \eqref{Vmatel}, results in $\mathrm{Tr}[ {\cal C}_{nn} \sigma_3 ] =0$. We obtain at this stage
\begin{equation}
M^{\mu\nu}_1 = -\tfrac{1}{2} \sum_n \left(N(\varepsilon_n)+\tfrac{1}{2}\right) \mathrm{Tr}\left(  {\cal C}_{nn} \right) 
\end{equation}


Simple calculation shows that 
\begin{equation}
\begin{aligned}
& i\, \mathrm{Tr} \left((\hat{V}^{\mu})_{nm}\sigma_3(\hat{V}^{\nu})_{mn} - (\hat{V}^{\nu})_{nm}\sigma_3(\hat{V}^{\mu})_{mn}\right) = 2\left[X_{a,nm}^{\mu\nu} + X_{b,nm}^{\mu\nu}\right] \,, 
\\
& i\, \mathrm{Tr} \left(\sigma_3(\hat{V}^{\mu})_{nm}(\hat{V}^{\nu})_{mn} - (\hat{V}^{\nu})_{nm}(\hat{V}^{\mu})_{mn}\sigma_3\right) = 2\left[X_{a,nm}^{\mu\nu} - X_{b,nm}^{\mu\nu}\right] \,,
\end{aligned}
\end{equation}
with combinations $X^{\mu\nu}_{c,nm}$ introduced in \eqref{X}. Finally, we obtain 
\begin{equation}
M^{\mu\nu}_1 = -\sum_{n,m} \left(N(\varepsilon_n) + \frac{1}{2}\right)\varepsilon_n\left[\frac{X_{a,nm}^{\mu\nu}}{\varepsilon_n - \varepsilon_m} + \frac{X_{b,nm}^{\mu\nu}}{\varepsilon_n + \varepsilon_m}\right] \,, 
\end{equation}
which is  Eq.\ \eqref{coefM} in the main text. It looks different from Eq. (D13) in  \cite{Matsumoto2014}. 
}

\end{widetext} 
 \end{document}